\documentclass[12p]{article}

\usepackage{amsmath}
\usepackage{amsfonts}
\usepackage{amssymb}
\usepackage{epsfig}
\usepackage{graphics}
\usepackage{subfig}

\usepackage[a4paper, margin=3cm]{geometry}

\newtheorem{proposition}{Proposition}[section]

\begin{document}

\title{ Exactly solvable variable parametric
 Burgers type models}
\maketitle

\begin{center}
{\bf \c{S}irin A. B\"{u}y\"{u}ka\c{s}{\i}k}\\
Izmir Institute of Technology, Mathematics Dept., \\
35430 Urla, Izmir, Turkey\\
sirinatilgan@iyte.edu.tr
\end{center}

\begin{abstract}
Exactly solvable variable parametric  Burgers type equations in one-dimension are introduced,
and two different approaches for solving the corresponding initial value problems are given.
The first one is using the relationship between  the variable parametric models and their standard counterparts. The second approach is a direct linearization of the variable parametric Burgers model to a variable parametric parabolic model via a generalized Cole-Hopf transform. Eventually, the problem of finding analytic and exact solutions of the variable parametric models reduces to that of solving a corresponding second order linear ODE with time dependent coefficients. This makes our results applicable to a wide class of exactly solvable Burgers type equations
 related with the classical Sturm-Liouville problems for the orthogonal polynomials.
\end{abstract}

\section{Introduction}

The nonlinear diffusion equation was known to Forsyth \cite{Forsyth}, and had been discussed by Bateman in connection with various viscous flows \cite{Bateman}.  J.M. Burgers  considered this equation as a model of turbulence \cite{Burgers1,Burgers}, Hopf  and Cole  discussed the Burgers equation in context of gas dynamics \cite{Hopf},\cite{Cole}, and Lighthill in acoustics \cite{Lighthill}. Recently, Burgers-KPZ turbulence was addressed in \cite{Woyczynski} and  \cite{Bec}, where the Burgers equation was considered also as a good approximation to understand the formation and distribution of matter at large scales.

Mathematically,  the standard Burgers equation (BE) $\,\, V_{t}+ V V_{x}=\nu V_{xx},\,\,$ is one of the best known integrable evolution equations, admitting  Lax representation, Painlev\'{e} property, B\"{a}cklund transformation, etc, \cite{Weiss}. It is a member of the infinite Burgers hierarchy of evolution equations which is linearizable in terms of the heat hierarchy, and analytic solutions can be easily found via the  Cole-Hopf transform. Various type of exact solutions  important in theory and applications are known for the BE. For example, when  the nonlinear effect and the effects of dissipative  nature  balance each other,  BE exhibits shock and multi-shock solitary wave solutions, which can fuse and therefore describe non-elastic interactions \cite{Whitham}, \cite{Wang}. On the other hand, it was shown that the motion of the poles of rational type solutions of BE corresponds to the motion of  classical particles interacting via simple two-body potentials. Then, generalization of this idea was used to discover new integrable many-body problems, see \cite{Choodnovsky}, \cite{Calogero}.

Recently, there is an increasing interest in nonlinear evolution equations with variable coefficients. In particular, for BE the variable coefficients are able to reflect  inhomogeneities in media, nonuniformities of boundaries, while the external forces, can model for example time evolution of the profile of growing interfaces, \cite{Xu}.  That makes the forced Burgers models with variable coefficients more realistic and significant for applications in various fields. In general, however, such models are not integrable and rarely admit exact solutions. For some positive results in this direction, one can see \cite{Xu}, \cite{Vallee},\cite{Eule},\cite{Axel},\cite{Zola},\cite{Sophocleous}.

The main subject of this work is the one-dimensional variable parametric Burgers equation of the form
\begin{equation}\label{burgers}
 U_{t}+\frac{\dot{\mu}(t)}{\mu(t)}U+U U_{x}=\frac{1}{2\mu(t)}U_{xx}-\omega^{2}(t)x\,,\,\,\,\,\,\,\,\,\,\,-\infty<x<\infty
\end{equation}
where $\Gamma(t)=\dot{\mu}(t)/\mu(t)$  is the damping term,  $D(t)=1/2\mu(t)$ is the diffusion coefficient, and $F(x,t)=-\omega^{2}(t)x$ is the forcing term which is linear in the space  variable $x.$

This model was introduced in our previous work \cite{sir-ok}, where exact solutions such as shock and multi-shock solitary waves, triangular waves, N-waves and rational type solutions were obtained. Here, we give an alternative way for finding these solutions and show the equivalence of both approaches. From this point of view, our results can be seen as complimentary to the previous ones. On the other side, this work is an extension which includes also the corresponding variable parametric potential Burgers and the linear parabolic equation. Thus, we provide a complete picture showing the relationships between the linear, the nonlinear, variable parametric and the standard models related with BE (\ref{burgers}).

In Sec.2, first we show that the  parabolic problem with variable coefficients and quadratic potential,  which appears as a linearization of BE (\ref{burgers}), can be reduced to that of solving  the classical heat equation and a corresponding linear ODE. Second, we solve the IVP for the variable parabolic model by finding explicitly the  evolution operator as a product of formal exponential operators. We remark that, the parabolic models discussed in this article are considered mainly as a linearization of the  Burgers type models. However,  solving some problems with specific initial conditions, we show that the results of Sec.2  can be used also independently  to discuss some heat or diffusion processes.

In Sec.3 and Sec.4 respectively, we consider the variable parametric potential Burgers and Burgers type models.  We obtain exact solutions of them using two different  approaches. The first one, is transforming the variable parametric Burgers type model to a standard  Burgers  equation, which in turn can be linearized in the form of the classical heat equation. This approach was applied in previous work \cite{sir-ok}, where different type of exact solutions were given and analyzed for models with constant damping and exponentially decaying diffusion coefficient.
The second approach is using a generalized Cole-Hopf transform, so that the variable parametric Burgers model is directly linearized in the form of the parabolic equation with time variable parameters discussed in Sec.2. The idea is similar to that which we used to construct exactly solvable Schr\"{o}dinger-Burgers models for complex velocity, which linearization takes the form of a Schr\"{o}dinger equation for harmonic oscillator with time variable parameters,  \cite{S.O},  \cite{S.O.E}. Finally, we apply our methods to solve some specific IVPs, and illustrate the formal procedure. In Sec.5, we give a scheme which summarizes the main results of the article.

\section{Variable parametric parabolic type equation}

 In this section, we consider  a variable parametric parabolic model
 \begin{eqnarray}\label{par-eq1}
\left\{
\begin{array}{ll}
 \displaystyle \frac{\partial \Phi}{\partial t}=\frac{1}{2\mu(t)}\frac{\partial^{2}
\Phi}{\partial x^2}+\frac{\mu(t)\omega^{2}(t)}{2}x^{2}\Phi,\\
\Phi(x,t)|_{t=t_{0}}=\Phi(x,t_{0}),\,\,\,\,\,-\infty<x<\infty,
 \end{array}
\right.
\end{eqnarray}
 related with the  IVP for a second-order linear differential equation of the form
\begin{equation}\label{clas-eq}
\ddot{r}+\frac{\dot{\mu}(t)}{\mu(t)}\dot{r}+\omega^{2}(t)r=0,\,\,\,r(t_{0})=r_{0}\neq
0,\,\,\,\dot{r}(t_{0})=0.
\end{equation}
   In Proposition \ref{prop1}, we show that one can solve the IVP (\ref{par-eq1}) in terms of solution to the classical heat equation and solution $r(t)$ of the IVP (\ref{clas-eq}).

   \begin{proposition}\label{prop1}
    The  IVP for the variable parametric parabolic equation (\ref{par-eq1})
has solution of the form
\begin{eqnarray}\label{vheat-sol}
\Phi(x,t)=\sqrt{\frac{r(t_{0})}{r(t)}}\times
\exp\left(-\frac{\mu(t)\dot{r}(t)}{2r(t)}x^2\right)\times \varphi(\eta(x,t),\tau(t)),
\end{eqnarray}
where $r(t)$ satisfies the IVP (\ref{clas-eq}), the auxiliary functions are
\begin{eqnarray}\label{tau}
\eta(x,t)=\frac{r(t_{0})}{r(t)}x\,;\,\,\,\,\,\,\,\,\,\,\,\,\tau(t)=
r^{2}(t_{0})\int^{t}\frac{d\xi}{\mu(\xi)r^{2}(\xi)},\,\,\,\,\,\,\tau(t_{0})=0,
\end{eqnarray}
and $\varphi(\eta,\tau)$ is solution of the IVP for the classical heat equation
\begin{eqnarray}\label{heat-ic11}
\left\{
\begin{array}{ll}
   \varphi_{\tau}=\frac{1}{2}\varphi_{\eta\eta}\,,\\
\varphi(\eta,0)=\Phi(\eta,t_{0}),\,\,\,\,\,-\infty<\eta<\infty.
\end{array}
\right.
\end{eqnarray}
\end{proposition}

\textbf{Proof:} Using the ansatz  $ \,\,\Phi(x,t)=\exp[(g(t)-\rho(t)x^2)/2]\,\varphi(e^{g(t)}x,\tau(t)),\,\,$
one can show that, if the auxiliary functions $\rho,\,\,\tau,\,\,g$ satisfy the nonlinear system of ordinary differential equations
\begin{eqnarray}\label{Ric}
\dot{\rho}+\frac{1}{\mu(t)}\rho^{2}+\mu(t)\omega^{2}(t)=0,\,\,\,\,\,\rho(t_{0})=0,
\end{eqnarray}
\begin{equation}\label{tau-1}
\dot{\tau}-\frac{ e^{2g}}{\mu(t)}=0,\,\,\,\,\,\tau(t_{0})=0,
\end{equation}
\begin{equation}\label{es}
\dot{g}+\frac{\rho}{\mu(t)}=0,\,\,\,\,\,g(t_{0})=0.
\end{equation}
 then the IVP (\ref{par-eq1})  reduces  to  IVP (\ref{heat-ic11}). Note that, Eq.(\ref{Ric}) is a nonlinear Riccati equation which can be linearized in the form of Eq.(\ref{clas-eq}), using
$\rho(t)=\mu(t)\dot{r}(t)/r(t).$  Then, to find $g(t)$ one needs to solve
$\,\,\dot{g}(t)+\dot{r}(t)/r(t)=0,\,\,\,g(t_{0})=0.\,\,$
For $r(t)\neq 0$, it has real-valued solution $g(t)=-\ln|r(t)/r(t_{0})|,$  but in this work we shall allow $g(t)$ to be complex-valued, that is $g(t)=-\ln(r(t)/r(t_{0})),$ which leads to more natural results for the Burgers type equations.
  Then, the system is easily solved, and we obtain the  auxiliary functions in terms of solution $r(t)$ to the IVP (\ref{clas-eq}), that is
\begin{eqnarray*}
\rho(t)=\mu(t)\frac{\dot{r}(t)}{r(t)}\,;\,\,\,\,\,\,\,\,\,\,\,\,\,\,\tau(t)=
r^{2}(t_{0})\int^{t}\frac{d\xi}{\mu(\xi)r^{2}(\xi)}\,,\,\,\,\tau(t_{0})=0\,;
\,\,\,\,\,\,\,\,\,\,\,\,\,\,g(t)=\ln(\frac{r(t_{0})}{r(t)})\,\,.
\end{eqnarray*}
  Writing these functions back in the ansatz, gives solution (\ref{vheat-sol}).

 $\Box$

Since the  IVP (\ref{heat-ic11}) for the classical heat equation has solution
 \begin{eqnarray}\label{heat-1}
 \varphi(\eta,\tau)=\frac{1}{\sqrt{2\pi\tau}}
 \int_{-\infty}^{\infty}\exp\left[-\frac{(\eta-y)^2}{2\tau}\right]\varphi(y,0)dy,
 \end{eqnarray}
  then according to Proposition \ref{prop1}, the solution of the IVP (\ref{par-eq1}) is found  in the form
 \begin{eqnarray}\label{heat-sol}
\Phi(x,t)=\sqrt{\frac{r(t_{0})}{r(t)}}\times
\exp\left[-\frac{\mu(t)\dot{r}(t)}{2r(t)}x^2\right]\times \frac{1}{\sqrt{2\pi\tau(t)}}
 \int_{-\infty}^{\infty}\exp\left[-\frac{(\eta(x,t)-y)^2}{2\tau(t)}\right]\Phi(y,t_{0})dy,
\end{eqnarray}
where $r(t)$ is solution of IVP (\ref{clas-eq}), and $\eta(x,t),\,\,\tau(t)$ are as defined in (\ref{tau}).
In case $\Phi(x,t_{0})=\delta(x),$ one obtains the  fundamental solution
\begin{eqnarray}\label{heat-fund}
\Phi(x,t)= \sqrt{\frac{r(t_{0})}{2\pi\tau(t)r(t)}}\times
 \exp\left[-\left(\frac{r^2(t_{0})}{2\tau(t)r^2(t)}+\frac{\mu(t)\dot{r}(t)}{2r(t)}\right)x^2\right].
\end{eqnarray}
Solution (\ref{heat-sol}) can be written also in the form
\begin{eqnarray}
\Phi(x,t)=\int_{-\infty}^{\infty}K_{t}(x,y)\Phi(y,t_{0})dy,
\end{eqnarray}
where the integral kernel is
\begin{eqnarray}\label{heat-kern}
K_{t}(x,y)=\sqrt{\frac{r(t_{0})}{2\pi\tau(t)r(t)}}\times
\exp\left[-\frac{\mu(t)\dot{r}(t)}{2r(t)}x^2\right]\times
 \exp\left[-\frac{(\eta(x,t)-y)^2}{2\tau(t)}\right].
\end{eqnarray}

   Using  Proposition \ref{prop1},  we solve two linear diffusion problems.

\textbf{Heat IVP-A.} First we consider the well-known constant coefficient heat equation with quadratic potential
\begin{eqnarray}\label{heat-ivp1}
\left\{
\begin{array}{ll}
 \displaystyle\frac{\partial \Phi}{\partial t}=\frac{1}{2}\frac{\partial^{2}
\Phi}{\partial x^2}-
\frac{\tilde{\omega}^{2}}{2}x^{2}\Phi,\\
\Phi(x,t)|_{t=0}=\Phi(x,0),\,\,\,\,\,-\infty<x<\infty,
 \end{array}
\right.
\end{eqnarray}
where $\mu(t)=1,$ $\omega^{2}(t)=-\tilde{\omega}^{2},$  $\, \tilde{\omega}>0.$  To find the solution and  integral kernel of this problem, there are many different methods in the literature, one can see for example Ref.\cite{Cycon}, p.286, and Ref.\cite{Berline}, p.153.
As an alternative, we use the approach in Proposition \ref{prop1}. The related IVP is,
$\,\,\ddot{r}-\tilde{\omega}^{2}r=0,\,\,\,r(0)=r_{0}\neq 0,\,\,\dot{r}(0)=0,$
and  the auxiliary functions are
$$r(t)=r_{0}\cosh(\tilde{\omega}t),\,\,\,\,\,\,\eta(x,t)=sech(\tilde{\omega}t)x,
\,\,\,\,\tau(t)=\tanh(\tilde{\omega}t)/\tilde{\omega}.$$
Then, according to (\ref{vheat-sol}),  solution of the IVP (\ref{heat-ivp1}) is found in terms of solution $\varphi$ of the classical heat equation as
\begin{eqnarray*}\label{heat-sol1}
\Phi(x,t)=\sqrt{\frac{1}{ \cosh(\tilde{\omega}t)}}\times \exp\left[-\frac{1}{2}\tilde{\omega}\tanh(\tilde{\omega}t)x^2\right]\times \varphi\left(sech(\tilde{\omega}t)x,\frac{\tanh(\tilde{\omega}t)}{\tilde{\omega}}\right).
\end{eqnarray*}
Using (\ref{heat-sol}), the general solution takes the form
\begin{eqnarray}\label{heat-sol1}
\Phi(x,t)=\sqrt{\frac{\tilde{\omega}}{2\pi \sinh(\tilde{\omega}t)}}\times \exp\left[-\frac{1}{2}\tilde{\omega}\tanh(\tilde{\omega}t)x^2\right]
\int_{-\infty}^{\infty}\exp\left[-\frac{[sech(\tilde{\omega}t)x-y]^2}{2\tanh(\tilde{\omega}t)/\tilde{\omega}}\right]
\Phi(y,0)dy,
\end{eqnarray}
and in case $\Phi(x,0)=\delta(x),$ one obtains the fundamental solution
\begin{eqnarray}\label{heat-fund1}
\Phi(x,t)=\sqrt{\frac{\tilde{\omega}}{2\pi \sinh(\tilde{\omega}t)}}\times \exp\left[-\frac{\tilde{\omega}}{2}\frac{x^2}{\tanh(\tilde{\omega}t)}\right].
\end{eqnarray}
Solution (\ref{heat-sol1}), can be written also in the form
\begin{eqnarray}
\Phi(x,t)=\int_{-\infty}^{\infty}K_{t}(x,y)\Phi(y,0)dy,
\end{eqnarray}
where
\begin{eqnarray}\label{heat-ker1}
K_{t}(x,y)=\sqrt{\frac{\tilde{\omega}}{2\pi \sinh(\tilde{\omega}t)}}\times
\exp\left[-\frac{\tilde{\omega}}{2}\left(\frac{(x^2+y^2)}
{\tanh(\tilde{\omega}t)}-\frac{2xy}{\sinh(\tilde{\omega}t)}\right)\right],
\end{eqnarray}
is the well-known heat kernel, or the integral kernel of the operator $\exp(-t\hat{H}),$ where
$$\hat{H}=-\frac{1}{2}\frac{\partial^2}{\partial x^2}+\frac{\tilde{\omega}^2}{2}x^2,\,\,\,\,\,\,on\,\,L_{2}(R).$$

\begin{figure}[h!]
\centering
\subfloat[ ]{\includegraphics[width=6.5cm,height=5.5cm]{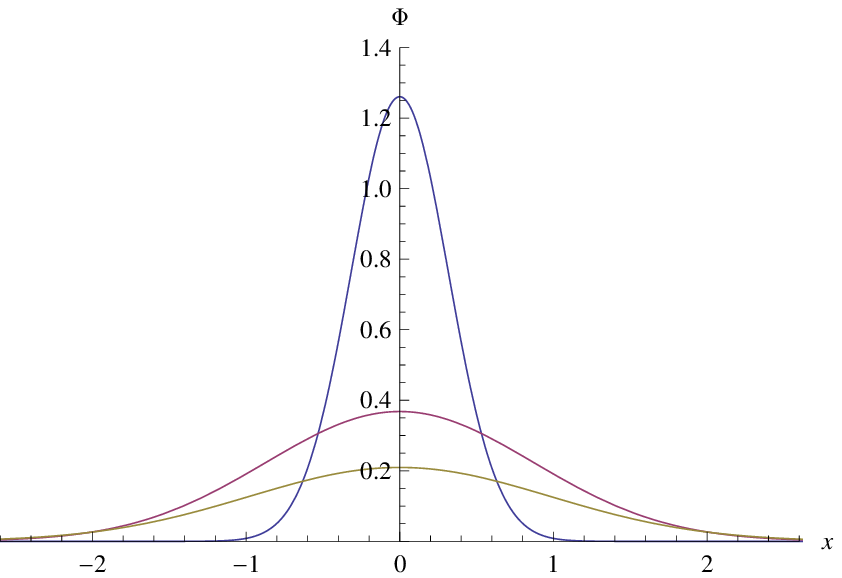}}
\hspace*{.51cm}
\subfloat[ ]{\includegraphics[width=6.5cm,height=5.5cm]{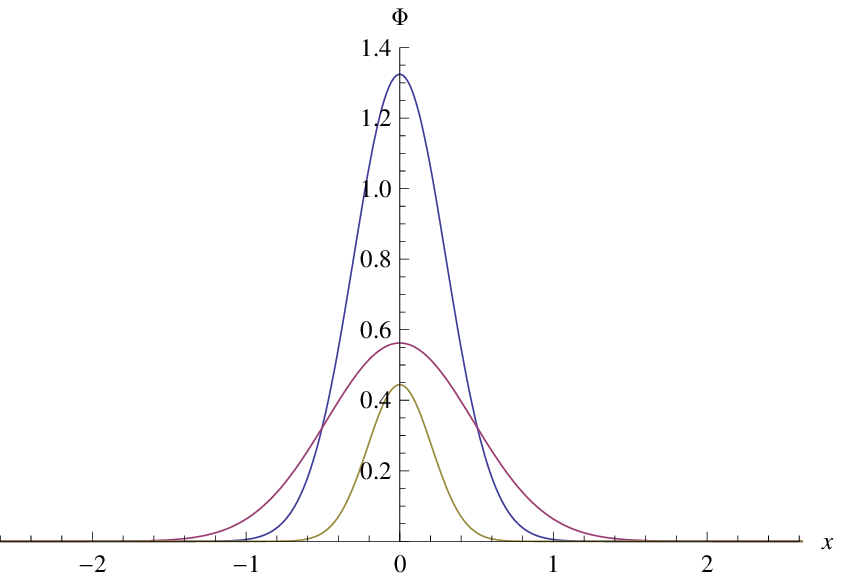}}
\vspace*{8pt}\caption{ a) Fundamental solution of Heat IVP-A at times $t=0.1, \,t=1,\,t=2,$ when $\tilde{\omega}=1.$\,\,\,\,\,\,\,\,\,
b) Fundamental solution of Heat IVP-B at times $t=0.1, \,t=1,\,t=2,$ when $\tilde{\omega}=1,$ \,$\gamma=2.$}
 \label{Heat-fund}
\end{figure}

\textbf{Heat IVP-B.} Consider the  heat equation with quadratic potential and time variable coefficients
\begin{eqnarray}\label{heat-ivp2}
\left\{
\begin{array}{ll}
 \displaystyle\frac{\partial \Phi}{\partial t}=\frac{1}{2}e^{-\gamma t}\frac{\partial^{2}
\Phi}{\partial x^2}-
\frac{\tilde{\omega}^{2}}{2}e^{\gamma t}x^{2}\Phi,\\
\Phi(x,t)|_{t=0}=\Phi(x,0),\,\,\,\,\,-\infty<x<\infty,
 \end{array}
\right.
\end{eqnarray}
where $\mu(t)=e^{\gamma t},\,\,\,\gamma>0,\,$ and $\omega^{2}(t)=-\tilde{\omega}^{2},$  $\, \tilde{\omega}>0.$  The corresponding IVP (\ref{clas-eq}) is
\begin{equation}\label{clas-eq-2}
\ddot{r}+\gamma \dot{r}-\tilde{\omega}^{2}r=0,\,\,r(0)=r_{0}\neq
0,\,\,\,\dot{r}(0)=0,
\end{equation}
 which has solution
\begin{eqnarray*}
r(t)=r_{0}\frac{\tilde{\omega}}{\tilde{\Omega}}e^{-\frac{\gamma t}{2}}\cosh[\tilde{\Omega}t+\tilde{\beta}],
\end{eqnarray*}
with $\tilde{\Omega}=\sqrt{\tilde{\omega}^{2} +(\gamma^2/4)},$ and $\tilde{\beta}=\tanh^{-1}(\gamma/2\tilde{\Omega}).$ Therefore,
$$\eta(x,t)=\frac{\tilde{\Omega}}{\tilde{\omega}}e^{\gamma t/2}\,sech[\tilde{\Omega}t+\tilde{\beta}],\,\,\,\,\,\,\,
\tau(t)=\frac{\tilde{\Omega}}{\tilde{\omega}^2}
\left(\tanh[\tilde{\Omega}t+\tilde{\beta}]-\frac{\gamma}{2\tilde{\Omega}}\right),$$
and according to (\ref{heat-sol}), the IVP (\ref{heat-ivp2}) has general solution
\begin{eqnarray}\label{heat-sol2}
\Phi(x,t)&=&\sqrt{\frac{\tilde{\Omega}e^{\gamma t/2}}{2\pi\tilde{\omega}\cosh[\tilde{\Omega}t+\tilde{\beta}]\tau(t)}}
\times \exp\left[-\frac{\tilde{\Omega}}{2}e^{\gamma t}\left(\tanh[\tilde{\Omega}t+\tilde{\beta}]-\frac{\gamma}{2\tilde{\Omega}}\right)x^2\right]\nonumber\\
& \times & \int_{-\infty}^{\infty}
\exp\left[-\frac{1}{2\tau(t)}\left(\frac{\tilde{\Omega}e^{\gamma t/2}x}{\tilde{\omega}\cosh[\tilde{\Omega}t+\tilde{\beta}]}
-y\right)^2 \right]\Phi(y,0)dy.
\end{eqnarray}
In particular, when the initial function is $\Phi(x,0)=\delta(x),$ we have the  fundamental solution
\begin{eqnarray*}
\Phi(x,t)&=&\sqrt{\frac{\tilde{\Omega}\,e^{\gamma t/2}}{2\pi\tilde{\omega}\cosh[\tilde{\Omega}t+\tilde{\beta}]\tau(t)}}
\times \exp\left[-\frac{\tilde{\Omega}}{2}e^{\gamma t}\left(\tanh[\tilde{\Omega}t+\tilde{\beta}]-\frac{\gamma}{2\tilde{\Omega}}\right)x^2\right]\nonumber\\
& \times & \exp\left[-\frac{\tilde{\Omega}^2}{2}e^{\gamma t}\left(\frac{1
}{\tilde{\omega}^2 \tau(t)\cosh^2[\tilde{\Omega}t+\tilde{\beta}]}\right)
 x^2\right],
\end{eqnarray*}
or
\begin{eqnarray}\label{heat-fund2}
\Phi(x,t)&=&\sqrt{\frac{\tilde{\Omega}\,e^{\gamma t/2}}{2\pi\tilde{\omega}\cosh[\tilde{\Omega}t+\tilde{\beta}]\tau(t)}}
\times \exp\left[-\frac{1}{2}e^{\gamma t}\left(\tilde{\omega}^{2}\tau(t)+
\frac{\tilde{\Omega}^2}{\tilde{\omega^2}\tau(t)\cosh^{2}[\tilde{\Omega}t+\tilde{\beta}]}\right)x^2\right].
\end{eqnarray}
 Note that, when $\gamma\rightarrow 0,$ one has $\tilde{\Omega}\rightarrow \tilde{\omega}$ and $\tilde{\beta}\rightarrow 0,$
so that solution (\ref{heat-fund2}) tends to  solution (\ref{heat-fund1}) of the constant coefficient Heat IVP-A.
The effect of the variable coefficients in IVP (\ref{heat-ivp2}), in particular the exponentially decreasing diffusion coefficient, can be seen by comparing Fig.\ref{Heat-fund}-a,  which shows  solution (\ref{heat-fund1}) and  Fig.\ref{Heat-fund}-b showing  solution (\ref{heat-fund2}).

\vspace{.2in}

Another approach to solve the IVP (\ref{par-eq1}) is  using the Evolution
Operator Method, also known as the Wei-Norman algebraic method,
\cite{Wei-Norman}, according to which, the evolution
operator can be represented as an ordered product of exponential
operators containing single generators of a Lie group. We obtain the following result.

\begin{proposition}\label{prop2}
Solution of the IVP (\ref{par-eq1}) is formally given by
\begin{equation}\label{wave2}
\Phi(x,t)=\hat{W}(t,t_{0})\Phi(x,t_{0}),
\end{equation}
where the evolution operator is
\begin{equation}\label{Wt-1}
\hat{W}(t,t_{0})=\exp\left(-\frac{\rho(t)}{2}x^{2}\right)\times\exp\left(
g(t)(x\frac{\partial}{\partial x}+\frac{1}{2})\right)\times
\exp\left(\frac{\tau(t)}{2}\frac{\partial^{2}}{\partial x^2}\right),
\end{equation}
and the auxiliary functions are found  in terms of solution $r(t)$ of the IVP (\ref{clas-eq}) as
\begin{eqnarray}\label{rho-tau-s}
\rho(t)=\mu(t)\frac{\dot{r}(t)}{ r(t)};\,\,\,\,\,\,\,\,\,\,\,\,
\tau(t)=r^{2}(t_{0})\int^{t}\frac{d\xi}{\mu(\xi)r^{2}(\xi)},\,\,\,\tau(t_{0})=0;\,\,\,\,\,\,\,\,\,\,
g(t)=-\ln\left(\frac{r(t)}{r(t_{0})}\right).
\end{eqnarray}
\end{proposition}

\noindent{\bf Proof:} The proof follows the same lines as for the Schr\"{o}dinger equation in paper Ref.\cite{S.O.E}. Indeed, for
the  IVP (\ref{par-eq1}),   the evolution
operator $\hat{W}(t,t_{0})$ can be found by solving the operator equation
\begin{equation}\label{Wt-2}
\frac{d}{d
t}\hat{W}(t,t_{0})=\hat{T}(t)\hat{W}(t,t_{0}),\,\,\,\,\,\,\,\,\,\hat{W}(t_{0},t_{0})=\hat{I},
\end{equation}
where
 \begin{equation}\hat{T}=\frac{1}{2\mu(t)}\frac{\partial^{2}
}{\partial
x^2}+\frac{\mu(t)\omega^{2}(t)}{2}x^{2}=i(\frac{1}{\mu(t)}\hat{K}_{-}-\mu(t)\omega^{2}(t)\hat{K}_{+}),
\end{equation}
and
\begin{equation}\label{SOP}
\hat{K}_{-}=-\frac{i}{2}\frac{\partial^{2}}{\partial
x^{2}},\,\,\,\,\,\,\,\phantom{a}\hat{K}_{+}=\frac{i}{2}x^{2},\,\,\,\,\,\,\,\phantom{a}
\hat{K}_{0}= \frac{1}{2}(x\frac{\partial}{\partial
x}+\frac{1}{2})\nonumber
\end{equation}
are operators which satisfy the commutation relations
\begin{equation}\label{SCR}
[\hat{K}_{-},\hat{K}_{+}]=2\hat{K}_{0},\,\,\,\,\phantom{a}
[\hat{K}_{+},\hat{K}_{0}]=-\hat{K}_{+},\,\,\,\,\phantom{a}
[\hat{K}_{-},\hat{K}_{0}]=\hat{K}_{-},\nonumber
\end{equation}
of the $SU(1,1)$ algebra. Assuming
\begin{equation}\label{Wt-3}
\hat{W}(t,t_{0})=\exp(i \rho(t)\hat{K}_{+})\times\exp(i\tau (t)
\hat{K}_{-})\times\exp(2g(t)\hat{K}_{0}),
\end{equation}
and taking time-derivative of $\hat{W},$ we get
\begin{eqnarray}\label{Wt-4}
\frac{d}{d t}\hat{W}(t,t_{0})=
\left[(i\dot{\rho}-i 2\rho\dot{g}-i \rho^{2}\dot{\tau}e^{-2g})\hat{K}_{+}
+(i\dot{\tau}e^{-2g})\hat{K}_{-}+(2\dot{g}+2\rho\dot{\tau}e^{-2g})\hat{K}_{0}\right]\hat{W}(t,t_{0}).
\end{eqnarray}
Comparing (\ref{Wt-2}) and (\ref{Wt-4}), it can be seen that
$\hat{W}(t,t_{0})$ is a solution of (\ref{Wt-2}), if the auxiliary
functions $\rho(t),$ $\tau(t)$ and $g(t)$ satisfy the non-linear system
of equations (\ref{Ric}),(\ref{tau-1}),(\ref{es}), which solution is given by (\ref{rho-tau-s}).
    Finally, since $\hat{W}$ satisfies (\ref{Wt-2}), it is well known that the function
$\Phi(x,t)=\hat{W}(t,t_{0})\Phi(x,t_{0})$ is a formal solution of the IVP
(\ref{par-eq1}). $\Box$

Combining the results of  Proposition \ref{prop1} and  Proposition \ref{prop2}, it follows that
  the evolution operator $\hat{W}(t,t_{0})$ given by (\ref{Wt-1}) has integral kernel $K_{t}(x,y)$ formally given by (\ref{heat-kern}), i.e.
$$\hat{W}(t,t_{0})\,\,.=\int_{-\infty}^{\infty}K_{t}(x,y)\,\,.\,\,dy.$$
 Moreover, using that
\begin{eqnarray*}
 \exp[\lambda x \frac{d}{dx}]f(x)=f(e^{\lambda}x)\,\,\,\,\,and\,\,\,\,\,\,     \exp[\frac{\tau(t)}{2}\frac{\partial^2}{\partial x^2}]\Phi(x,t_{0})=\varphi(x,\tau(t)),
\end{eqnarray*}
where $\varphi$ is  solution of the classical heat IVP (\ref{heat-ic11}), we have
\begin{eqnarray}\label{heat-sol-3}
\Phi(x,t)&=&\hat{W}(t,t_{0})\Phi(x,t_{0})\nonumber \\&=&\exp\left(-\frac{\rho(t)}{2}x^{2}\right)\times\exp\left(
g(t)(x\frac{\partial}{\partial x}+\frac{1}{2})\right)\times
\exp\left(\frac{\tau(t)}{2}\frac{\partial^{2}}{\partial x^2}\right)\Phi(x,t_{0})\nonumber\\
&=&\exp(\frac{1}{2}g(t))\times\exp(-\frac{\rho(t)}{2}x^2)\times \varphi(e^{g(t)}x,\tau(t)),
\end{eqnarray}
and writing the functions $\rho(t),\,\tau(t)$ and $g(t)$ as found in (\ref{rho-tau-s}), one can see that
solution (\ref{heat-sol-3}) obtained by the evolution operator method coincides with the previously found solution (\ref{vheat-sol}). This establishes the equivalence of both approaches.

Proposition \ref{prop2}  can be easily applied when the initial function is given in the form of a power series, or in particular if it is a linear combinations of exponential functions such as $\cosh$ and $\sinh,$
using the well known relations, see \cite{Dattoli},
\begin{equation}\label{aux}
\exp[\lambda \frac{d^{2}}{dx^2}]x^{m}=H_{m}(x,\lambda),\,\,\,\,\,\,\,\,\exp[\lambda
\frac{d^{2}}{dx^2}]e^{x}=e^{\lambda}e^{x},
\end{equation}
where $H_{m}(x,\lambda)$ are Kamp\`{e} de Feriet polynomials defined by
$$H_{m}(x,\lambda)=m!\sum_{k=0}^{[m/2]}\frac{\lambda^k}{k!(m-2k)!}x^{m-2k},\,\,\,\,\,H_{m}(x,0)=x^{m}.$$

Therefore, using Proposition \ref{prop2}, exact solutions of the following parabolic IVP's are found. We remark that,
these IVP's may not have direct physical meaning, but will be useful in Sec.4, to obtain special exact solutions for the corresponding variable parametric Burgers type models.

\textbf{IVP 2.1.} The IVP (\ref{par-eq1}) with initial condition $\Phi(x,t_{0})=e^{-cx}\cosh[Ax+c_{0}],\,\,\,\,\,-\infty<x<\infty,$ and
$A,\,c,\,c_{0}$  real constants,
 has solution
\begin{eqnarray}\label{sol-3}
\Phi(x,t)& = &\hat{W}(t,t_{0})\left(\frac{e^{(A-c)x+c_{0}}+e^{-(A+c)x-c_{0}}}{2}\right)\nonumber\\
& = & \sqrt{\frac{r(t_{0})}{r(t)}}\times \exp\left[-\frac{\mu(t)\dot{r}(t)}{2r(t)}x^2\right]\times \exp\left[ (\frac{A^2+c^2}{2})\tau(t)\right]\times\exp\left[-c\frac{r(t_{0})}{r(t)}x\right]\nonumber\\ & &\times \cosh\left[A (\frac{r(t_{0})}{r(t)}x-c\tau(t))+c_{0}\right],
\end{eqnarray}
where  $r(t)$ is solution of IVP (\ref{clas-eq}), $\tau(t)$ is as given in (\ref{rho-tau-s}).

\textbf{ IVP 2.2.} The IVP (\ref{par-eq1}) with initial condition
$\Phi(x,t_{0})=x^m,\,\,\,\,\,-\infty<x<\infty,\,\,\,\,\,\,m=0,1,2,...,$
has solution of the form
\begin{eqnarray}\label{sol-1}
\Phi_{m}(x,t)&=&\hat{W}(t,t_{0})x^{m}=\exp\left[\frac{1}{2}g(t)-\frac{1}{2}\rho(t)x^{2}\right]\times
H_{m}\left(e^{g(t)}x,\frac{1}{2}\tau(t)\right)\nonumber\\
& = & \sqrt{\frac{r(t_{0})}{r(t)}}\times \exp\left[-\frac{\mu(t)\dot{r}(t)}{2r(t)}x^2\right]\times
H_{m}\left(\frac{r(t_{0})}{r(t)}x,\,\frac{1}{2}\tau(t)\right),\,\,\,\,\,m=0,1,2,....
\end{eqnarray}

\vspace{.3in}

\textbf{General IVP 2.3 } The IVP (\ref{par-eq1}) with general initial condition
$\Phi(x,t_{0})=\sum_{m=0}^{\infty}a_{m}x^{m},$
 has formal solution
\begin{eqnarray}
\Phi(x,t)&=&\sum_{m=0}^{\infty}a_{m}\Phi_{m}(x,t)\nonumber\\
&=&\sqrt{\frac{r(t_{0})}{r(t)}}\times \exp\left[-\frac{\mu(t)\dot{r}(t)}{2r(t)}x^2\right]\times\sum_{m=0}^{\infty}a_{m}
H_{m}\left(\frac{r(t_{0})}{r(t)}x,\,\frac{1}{2}\tau(t)\right),
\end{eqnarray}
where $\Phi_{m}(x,t)$ is given by (\ref{sol-1}).

\section{ Variable parametric potential Burgers equation}

Now we consider the initial value problem for a variable parametric potential Burgers equation of the form
\begin{eqnarray}\label{pot-burgers-ic}
\left\{
\begin{array}{ll}
 \displaystyle\frac{\partial \Upsilon}{\partial t}+\frac{\dot{\mu}(t)}{\mu(t)}\Upsilon+\frac{1}{2}(\frac{\partial
\Upsilon}{\partial
x})^{2}=\frac{1}{2\mu(t)}\frac{\partial^{2}\Upsilon}{\partial
x^2}-\frac{\omega^{2}(t)}{2}x^{2} \,\,,\\
\Upsilon(x,t)|_{t=t_{0}}=\Upsilon(x,t_{0})\,.\,
\end{array}
\right.
\end{eqnarray}

 We note that, if we write  $\Phi(x,t)=\exp(-\mu(t)\Upsilon(x,t)),$  then the variable parametric parabolic equation  (\ref{par-eq1}) transforms formally to the variable parametric potential Burgers equation  (\ref{pot-burgers-ic}). On the other hand,  writing $\varphi(\eta,\tau)=\exp(-h(\eta,\tau))$ the standard heat equation transforms to the standard potential Burgers equation. Using this relations and Proposition \ref{prop1}, we formulate the following propositions.
\begin{proposition}\label{prop3} The IVP  (\ref{pot-burgers-ic}) for the variable parametric potential Burgers
equation has solution of the following forms:

a)  \begin{equation}
\Upsilon(x,t)=-\frac{1}{2\mu(t)}\ln\left(\frac{r(t_{0})}{r(t)}\right)+ \frac{1}{2}\frac{\dot{r}(t)}{r(t)}x^{2}+
\frac{1}{\mu(t)}h(\eta(x,t),\tau(t))
\end{equation}
where $r(t)$ is solution of IVP (\ref{clas-eq}),
\begin{eqnarray}
\eta(x,t)=\frac{r(t_{0})}{r(t)}x\,;\,\,\,\,\,\,\,\,\,\,\,\,\tau(t)=
r^{2}(t_{0})\int^{t}\frac{d\xi}{\mu(\xi)r^{2}(\xi)},\,\,\,\,\,\,\tau(t_{0})=0,
\end{eqnarray}
and $h(\eta,\tau)$ satisfies the IVP for the standard potential Burgers equation
\begin{eqnarray}\label{pot-burg-con}
\left\{
\begin{array}{ll}
 \displaystyle h_{\tau}+\frac{1}{2}h_{\eta}^{2}=\frac{1}{2}h_{\eta\eta}
 \,\,,\\
h(\eta,0)=\mu(t_{0})\Upsilon(\eta,t_{0})
\end{array}
\right.
\end{eqnarray}

b) \begin{equation}
\Upsilon(x,t)=-\frac{1}{2\mu(t)}\ln\left(\frac{r(t_{0})}{r(t)}\right)+ \frac{1}{2}\frac{\dot{r}(t)}{r(t)}x^{2}-
\frac{1}{\mu(t)}\ln\varphi(\eta(x,t),\tau(t))
\end{equation}
where $\eta,\tau$ are as defined in part (a), and $\varphi(\eta,\tau)$ satisfies the IVP for the heat equation
\begin{eqnarray}\label{heat-ic1}
\left\{
\begin{array}{ll}
   \varphi_{\tau}=\frac{1}{2}\varphi_{\eta\eta}\,,\\
\varphi(\eta,0)=\exp[-\mu(t_{0})\Upsilon(\eta,t_{0})].\,\,\,\Box
\end{array}
\right.
\end{eqnarray}
\end{proposition}

Using the general solution (\ref{heat-1}) of the heat equation and the relation $h(\eta,\tau)=-\ln\varphi(\eta,\tau),$ one can easily write the analytic solution of IVP (\ref{pot-burg-con}), that is
\begin{eqnarray}\label{pot-bur-1}
 h(\eta,\tau)=-\ln\left[\frac{1}{\sqrt{2\pi\tau}}
 \int_{-\infty}^{\infty}\exp\left[-\mu(t_{0})\Upsilon(y,t_{0})-\frac{(\eta-y)^2}{2\tau}\right]dy\right],
 \end{eqnarray}
 As a result, the general solution of  IVP (\ref{pot-burgers-ic}) is found in the form
 \begin{eqnarray}
\Upsilon(x,t)&=&-\frac{1}{2\mu(t)}\ln\left(\frac{r(t_{0})}{r(t)}\right)+ \frac{1}{2}\frac{\dot{r}(t)}{r(t)}x^{2}\nonumber\\
& & -\frac{1}{\mu(t)}\ln\left[\frac{1}{\sqrt{2\pi\tau(t)}}
 \int_{-\infty}^{\infty}\exp\left[-\mu(t_{0})\Upsilon(y,t_{0})-\frac{(\eta(x,t)-y)^2}{2\tau(t)}\right]dy\right].
\end{eqnarray}

Next proposition gives direct relation of the variable parametric potential Burgers IVP with the variable parametric parabolic IVP.
\begin{proposition}\label{prop4}
The IVP (\ref{pot-burgers-ic}) for the variable parametric potential Burgers equation
has  formal solution
\begin{equation}
\Upsilon(x,t)=-\frac{1}{\mu(t)}\ln\Phi(x,t),
\end{equation}
where $\Phi(x,t)$ satisfies
the IVP for the variable parametric parabolic equation
\begin{eqnarray}\label{par-eq-1}
\left\{
\begin{array}{ll}
 \displaystyle \frac{\partial \Phi}{\partial t}=\frac{1}{2\mu(t)}\frac{\partial^{2}
\Phi}{\partial x^2}+\frac{\mu(t)\omega^{2}(t)}{2}x^{2}\Phi,\,\,,\\
\Phi(x,t_{0})=\exp(-\mu(t_{0})\Upsilon(x,t_{0})).\,\,\,\,\Box
\end{array}
\right.
\end{eqnarray}
\end{proposition}

\section{ Variable parametric Burgers equation}

Finally, we discuss the IVP for a one-dimensional variable parametric Burgers equation
\begin{eqnarray}\label{burgers-ic1}
\left\{
\begin{array}{ll}
  \displaystyle \frac{\partial U}{\partial t}+\frac{\dot{\mu}(t)}{\mu(t)}U+U\frac{\partial U}{\partial
x}=\frac{1}{2\mu(t)}\frac{\partial^{2}U}{\partial
x^2}-\omega^{2}(t)x\,\,,\\
U(x,t)|_{t=t_{0}}=U(x,t_{0})\,\,,\,\,\,\,-\infty<x<\infty
\end{array}
\right.
\end{eqnarray}
where $\Gamma(t)=\dot{\mu}(t)/\mu(t)$  is the damping term,  $D(t)=1/2\mu(t)$ is the diffusion coefficient, and $F(x,t)=-\omega^{2}(t)x$ is the forcing term which is linear in the space  variable $x.$

  In \cite{sir-ok}, it was shown that solutions of the IVP (\ref{burgers-ic1}) can be obtained in terms of solutions to the standard Burgers equation.  For completeness and comparison, in Proposition \ref{prop5} we outline this approach. Using it, special exact solutions such as generalized shock and multi-shock solitary waves, triangular waves, \emph{N}-waves  and rational type solutions were found  and discussed for forced Burgers equations with constant damping and exponentially decaying diffusion coefficients.

\begin{proposition}\label{prop5} The IVP for the variable parametric Burgers equation (\ref{burgers-ic1})
has solution in the following forms:
\begin{eqnarray}\label{sol-U}
a)\,\,\,\,\,\,\,\,\,\,\,\,\,\,\,\,\,\,\,\,\,\,\,\,\,\,\,\,\,\,\,\,\,\,\,\,
U(x,t)=\frac{\dot{r}(t)}{r(t)}x+\frac{r(t_{0})}{\mu(t)r(t)}V\left(\eta(x,t),\tau(t)\right),\,
\end{eqnarray}
where $r(t)$ is a solution of IVP (\ref{clas-eq}),
\begin{equation}\label{eta-tau}
\eta(x,t)=\frac{r(t_{0})}{r(t)}x;\,\,\,\,
\,\,\,\,\,\tau(t)=r^{2}(t_{0})\int^{t}\frac{d\xi}{\mu(\xi)r^{2}(\xi)}\,\,,\,\,\tau(t_{0})=0,
\end{equation}
and the function $V(\eta,\tau)$ satisfies the IVP for the standard Burgers equation
\begin{eqnarray}\label{HBE-ic1}
\left\{
\begin{array}{ll}
  V_{\tau}+V V_{\eta}=\frac{1}{2}V_{\eta\eta},\,\\
V(\eta,0)=\mu(t_{0})U(\eta,t_{0})
\end{array}
\right.
\end{eqnarray}

\begin{eqnarray}\label{sol-U2}
b)\,\,\,\,\,\,\,\,\,\,\,\,\,\,\,\,\,\,\,\,\,\,\,\,\,\,\,\,\,\,\,\,\,\,\,\,
U(x,t)=\frac{\dot{r}(t)}{r(t)}x-
\frac{r(t_{0})}{\mu(t)r(t)}\frac{\varphi_{\eta}(\eta(x,t),\tau(t))}{\varphi(\eta(x,t),\tau(t))},
\end{eqnarray}
where $\eta,$ $\tau$ are as defined in part (a), and $\varphi(\eta,\tau)$ satisfies the IVP for the heat equation
 \begin{eqnarray}\label{heat-ic2}
\left\{
\begin{array}{ll}
   \varphi_{\tau}=\frac{1}{2}\varphi_{\eta\eta}\,,\\
\varphi(\eta,0)=\exp\left[-\int^{\eta}\mu(t_{0})U(\xi,t_{0})d\xi\right]. \,\,\,\,\Box
\end{array}
\right.
\end{eqnarray}
\end{proposition}

The well known solution (\ref{heat-1}) of the IVP (\ref{heat-ic2})
and the Cole-Hopf transformation $V=-\varphi_{\eta}/\varphi,$ lead to solution of the IVP (\ref{HBE-ic1}) for the Burgers equation
\begin{eqnarray*}
V(\eta,\tau)=\frac{\int_{-\infty}^{\infty}
\left(\frac{\eta-y}{\tau}\right)\exp\left[-\left(\frac{(\eta-y)^2}{2\tau}+\int^{y}V(y ' ,0)d y ' \right)\right]dy}
{\int_{-\infty}^{\infty}
\exp\left[-\left(\frac{(\eta-y)^2}{2\tau}+\int^{y}V(y ' ,0)dy ' \right)\right]dy}.
\end{eqnarray*}
Therefore, using  Proposition \ref{prop5}, one can find formal solution of the IVP (\ref{burgers-ic1}) for the variable parametric Burgers equation in terms of solution $r(t)$ of the IVP (\ref{clas-eq}), that is
\begin{eqnarray*}\label{gen-sol}
U(x,t)=\frac{\dot{r}(t)}{r(t)}x \hspace{5in}\\
+\left[\frac{r(t_{0})}{\mu(t)r(t)}\right]\frac{\int_{-\infty}^{\infty}
\left(\frac{1}{\tau(t)}(\frac{r(t_{0})}{r(t)}x-y)\right)
\exp\left[-\left(\frac{1}{2\tau(t)}(\frac{r(t_{0})}{r(t)}x-y)^2+\int^{y}\mu(t_{0})U(y ' ,t_{0})d y ' \right)\right]dy}
{\int_{-\infty}^{\infty}
\exp\left[-\left(\frac{1}{2\tau(t)}(\frac{r(t_{0})}{r(t)}x-y)^2+\int^{y}\mu(t_{0})U(y ' ,t_{0})d ' \right)\right]dy}.
\end{eqnarray*}

\vspace{.2in}

Next proposition establishes direct relation between the variable parametric BE and the variable parametric parabolic equation via a generalized Cole-Hopf transformation.  This gives us an alternative way of solving the IVP (\ref{burgers-ic1}).

\begin{proposition}\label{prop6}
   The IVP (\ref{burgers-ic1}) for the variable parametric Burgers equation
has solution of the  forms:
 \begin{equation}
a) \,\,\,\,\,\,\,\,\,\,\,\,\,\,\,\,\,\,\,\,\,U(x,t)=\frac{\partial}{\partial x}\Upsilon(x,t),
\end{equation}
where $\Upsilon(x,t)$ is  solution of the IVP for the variable parametric potential BE
\begin{eqnarray}\label{pot-burgers-ic1}
\left\{
\begin{array}{ll}
 \displaystyle\frac{\partial \Upsilon}{\partial t}+\frac{\dot{\mu}(t)}{\mu(t)}\Upsilon+\frac{1}{2}(\frac{\partial
\Upsilon}{\partial
x})^{2}=\frac{1}{2\mu(t)}\frac{\partial^{2}\Upsilon}{\partial
x^2}-\frac{\omega^{2}(t)}{2}x^{2} \,\,,\\
\Upsilon(x,t_{0})=\int^{x}U(\xi,t_{0})d\xi\,.\,\,\,\,\,\,
\end{array}
\right.
\end{eqnarray}

\begin{equation}
b)\,\,\,\,\,\,\,\,\,\,\,\,\,\, U(x,t)=-\frac{1}{\mu(t)}\frac{\partial}{\partial x}(\ln\Phi(x,t)),
\end{equation}
where $\Phi(x,t)$ satisfies the IVP for the variable parametric parabolic equation
\begin{eqnarray}\label{par-eq-1}
\left\{
\begin{array}{ll}
 \displaystyle \frac{\partial \Phi}{\partial t}=\frac{1}{2\mu(t)}\frac{\partial^{2}
\Phi}{\partial x^2}+\frac{\mu(t)\omega^{2}(t)}{2}x^{2}\Phi,\,\,\\
\Phi(x,t_{0})=\exp\left(-\mu(t_{0})\int^{x}U(\xi,t_{0})d\xi\right).\,\,\,\,\Box
\end{array}
\right.
\end{eqnarray}
\end{proposition}

 Proof can be done by direct calculation. In what follows, results of Sec.2 are combined with Proposition \ref{prop6}-(b) to obtain analytic and exact solutions of some Burgers IVPs.

\textbf{Burgers IVP-A.} Given the Burgers problem with linear external potential
 \begin{eqnarray}\label{Burg-IVP-A}
\left\{
\begin{array}{ll}
 \displaystyle
   U_{t}+U U_{x}=\frac{1}{2}U_{xx}+\tilde{\omega}^{2}x,\\
   U(x,t)|_{t=0}=U(x,0),\,\,\,\,\,\,-\infty<x<\infty,
   \end{array}
\right.
\end{eqnarray}
where  $\tilde{\omega}>0,$ it is not difficult to see that its linearization  takes the form of the Heat IVP-A. Therefore, its solution is  $U(x,t)=-\partial_{x}(\ln \Psi(x,t)),$ where $\Psi(x,t)$ is given by (\ref{heat-sol1}). Explicitly one has
\begin{eqnarray*}\label{gen-sol}
U(x,t)=\tilde{\omega}\tanh(\tilde{\omega}t)x+sech(\tilde{\omega}t)\frac{\int_{-\infty}^{\infty}
\left(\frac{sech(\tilde{\omega}t)x-y}{\tanh(\tilde{\omega}t)/\tilde{\omega}}\right)
\exp\left[-\left(\frac{sech(\tilde{\omega}t)x-y)^2}{2\tanh(\tilde{\omega}t)/\tilde{\omega}}+\int^{y}U(\xi,0)d \xi  \right)\right]dy}
{\int_{-\infty}^{\infty}
\exp\left[-\left(\frac{sech(\tilde{\omega}t)x-y)^2}{2\tanh(\tilde{\omega}t)/\tilde{\omega}}+\int^{y}U(\xi,0)d\xi \right)\right]dy},
\end{eqnarray*}
which is a known solution, Refs.\cite{Leonenko1},\cite{Leonenko2}.

\textbf{Burgers IVP-B.} The forced Burgers problem with damping term and exponentially decaying diffusion coefficient
\begin{eqnarray}\label{Burg-IVP-A}
\left\{
\begin{array}{ll}
 \displaystyle
   U_{t}+\gamma U+U U_{x}=\frac{1}{2}e^{-\gamma t}U_{xx}+\tilde{\omega}^{2}x,\,\,\,\,\,\, \tilde{\omega}>0, \\
   U(x,t)|_{t=0}=U(x,0),\,\,\,\,\,\,-\infty<x<\infty,
   \end{array}
\right.
\end{eqnarray}
 is linearazible in the form of Heat IVP-B. Therefore, its solution is
$U(x,t)=-e^{-\gamma t}\partial_{x}(\ln \Psi(x,t)),$ where $\Psi(x,t)$ is given by (\ref{heat-sol2}). Explicitly one has
\begin{eqnarray*}
U(x,t)=\left(-\frac{\gamma}{2}+\tilde{\Omega}\tanh[\tilde{\Omega}t+\tilde{\beta}]\right)x \hspace{3.5in}\\
+\frac{\tilde{\Omega}e^{-\gamma t/2}}{\tilde{\omega}\cosh[\tilde{\Omega}t+\tilde{\beta}]\,\tau(t)}\int_{-\infty}^{\infty}\left(\frac{\tilde{\Omega}e^{\gamma t/2}x}{\tilde{\omega}\cosh[\tilde{\Omega}t+\tilde{\beta}]}
-y\right)
\exp\left[-\frac{1}{2\tau(t)}\left(\frac{\tilde{\Omega}e^{\gamma t/2}x}{\tilde{\omega}\cosh[\tilde{\Omega}t+\tilde{\beta}]}
-y\right)^2 \right]\Phi(y,0)dy\\
 \times \left(\int_{-\infty}^{\infty}
\exp\left[-\frac{1}{2\tau(t)}\left(\frac{\tilde{\Omega}e^{\gamma t/2}x}{\tilde{\omega}\cosh[\tilde{\Omega}t+\tilde{\beta}]}
-y\right)^2 \right]\Phi(y,0)dy\right)^{-1}.
\end{eqnarray*}

\textbf{IVP 4.1.} The IVP (\ref{burgers-ic1}) with initial condition
$$U(x,t_{0})
=\frac{1}{\mu(t_{0})}\left(c-
A\tanh\left[A x+c_{0}\right]\right)\,,\,\,\,\,\,-\infty<x<\infty,$$
has   single shock solitary type solution
 \begin{eqnarray}\label{U-tanh-g}
U(x,t)&=&-\frac{1}{\mu(t)}\frac{\partial}{\partial x}(\ln\Phi(x,t))\\
&=&\frac{\dot{r}(t)}{r(t)}x+\frac{r(t_{0})}{\mu(t)r(t)}\left(c-
A\tanh\left[A\left( \frac{r(t_{0})}{r(t)}x-c\tau(t)\right)+c_{0}\right]\right).
\end{eqnarray}
Here we used that $\Phi(x,t)$ is a solution of the parabolic IVP 2.1 given in Sec.2.

\textbf{ IVP 4.2.} The IVP (\ref{burgers-ic1}) with initial conditions
$$U_{m}(x,t_{0})=-\frac{m}{\mu(t_{0})}\frac{1}{x}\,\,,\,\,\,\,m=0,1,2,...$$
has rational type solutions
\begin{eqnarray}
U_{m}(x,t)&=&-\frac{1}{\mu(t)}\frac{\partial}{\partial x}(\ln\Phi_{m}(x,t))=\frac{1}{\mu(t)}\left[\rho(t)x-\frac{\partial }{\partial x}
\left(\ln H_{m}\left(e^{g(t)}x,\frac{1}{2}\tau(t)\right)\right)\right]\nonumber\\
&=&\frac{\dot{r}(t)}{r(t)}x-\frac{m}{\mu(t)} (\frac{r(t_{0})}{r(t)})\left(\frac{H_{m-1}\left(\frac{r(t_{0})}{r(t)}x,\frac{1}{2}\tau(t)\right)}
{H_{m}\left(\frac{r(t_{0})}{r(t)}x,\frac{1}{2}\tau(t)\right)}\right),\,\,\,\,\,m=0,1,2,...
\end{eqnarray}
where we used that $\Phi_{m}(x,t)$ is a solution of the parabolic IVP 2.2, and $\frac{\partial}{\partial x}H_{m}(x,\lambda)=m H_{m-1}(x,\lambda).$

\textbf{General  IVP 4.3.} The IVP (\ref{burgers-ic1}) with the general initial condition
$$U(x,t)=-\frac{1}{\mu(t_{0})}\frac{\partial}{\partial x}\left(\ln\sum_{n=0}^{\infty}a_{m}x^{m}\right),$$
has formal solution given by
$$U(x,t)=-\frac{1}{\mu(t)}\frac{\partial}{\partial x}(\ln\Phi(x,t))
=\frac{\dot{r}(t)}{r(t)}x-\frac{1}{\mu(t)}\frac{\partial}{\partial x}\left(\ln\sum_{m=0}^{\infty}a_{m}
H_{m}\left(\frac{r(t_{0})}{r(t)}x,\,\frac{1}{2}\tau(t)\right)\right),$$
where we used that $\Phi(x,t)$ is a solution of the parabolic IVP 2.3.

We note that the above special IVP's were discussed in \cite{sir-ok}, but using Proposition \ref{prop5}. Here, we illustrate the application of Proposition \ref{prop6}-(b), and show that both approaches lead to the same results.

\section{Summary}

In this work,  variable parametric Burgers type models are introduced
and two different approaches for solving the corresponding initial value problems are given.
The first approach is transforming the variable parametric model to a standard constant coefficient model. The second approach is a direct linearization of the variable parametric Burgers model to a variable parametric parabolic model.
Both approaches and the relations between them  are summarized in the following scheme.
\begin{eqnarray*}
\begin{array}{ccc}
\underline{Variable\,\, Parametric\,\, Models }&  & \underline{Standard\,\, Models}\\
\Downarrow &  & \Downarrow\\
 \displaystyle \boxed{ U_{t}+\frac{\dot{\mu}(t)}{\mu(t)}U+U U_{x}=\frac{1}{2\mu(t)}U_{xx}-\omega^{2}(t)x} & \longrightarrow & \boxed{V_{\tau}+V V_{\eta}=\frac{1}{2}V_{\eta\eta}} \\
  \downarrow & & \downarrow\\
  U(x,t)=\Upsilon_{x}(x,t) & & V(\eta,\tau)= h_{\eta}(\eta,\tau)\\
  \downarrow & & \downarrow\\
 \displaystyle \boxed{ \Upsilon_{t}+\frac{\dot{\mu}(t)}{\mu(t)}\Upsilon+\frac{1}{2}(
\Upsilon_{
x})^{2}=\frac{1}{2\mu(t)}\Upsilon_{xx}-\frac{\omega^{2}(t)}{2}x^{2}}& \longrightarrow &  \boxed{h_{\tau}+\frac{1}{2}h_{\eta}^{2}=\frac{1}{2}h_{\eta\eta}}\\
\downarrow & & \downarrow\\
\Upsilon(x,t)=-\frac{1}{\mu(t)}\ln\Phi(x,t) &  & h(\eta,\tau)=-\ln\varphi(\eta,\tau)\\
\downarrow & & \downarrow\\
\displaystyle\boxed{ \Phi_{t}=\frac{1}{2\mu(t)}
\Phi_{xx}+\frac{\mu(t)\omega^{2}(t)}{2}x^{2}\Phi} & \longrightarrow& \boxed{\varphi_{\tau}=\frac{1}{2}\varphi_{\eta\eta}}
\end{array}
\end{eqnarray*}

At a final stage, the problem of solving the variable parametric models introduced in this article, reduces to that of finding solution of a corresponding second order linear ODE with time dependent coefficients.
This will allow us to study a wide class of exactly solvable Burgers type models related with the classical Sturm-Liouville problems for the  orthogonal polynomials, or special functions.

\vspace{.2in}

\textbf{Acknowledgments:} This work is supported by the National Science Foundation of Turkey, T\"{U}BITAK, TBAG Project No: 110T679.

\end{document}